\documentclass[twocolumn,superscriptaddress,showpacs,preprintnumbers,amsmath,amssymb]{revtex4}

\usepackage{graphicx}
\usepackage{dcolumn}
\usepackage{bm}
\usepackage{color}
\usepackage{soul}
\usepackage{ulem}

\begin{document}

\preprint {accepted manuscript for Phys. Rev. Appl.}

\title{Interface spin polarization of the Heusler compound Co$_2$MnSi probed by unidirectional spin Hall magnetoresistance}

\author{C. Lidig}
\affiliation{Institut f\"ur Physik, Johannes Gutenberg-Universit\"at, Staudinger Weg 7, 55128 Mainz, Germany}
\author{J. Cramer}
\affiliation{Institut f\"ur Physik, Johannes Gutenberg-Universit\"at, Staudinger Weg 7, 55128 Mainz, Germany}
\author{L. Wei{\ss}hoff}
\affiliation{Institut f\"ur Physik, Johannes Gutenberg-Universit\"at, Staudinger Weg 7, 55128 Mainz, Germany}
\author{T. R. Thomas}
\affiliation{Institut f\"ur Physik, Johannes Gutenberg-Universit\"at, Staudinger Weg 7, 55128 Mainz, Germany}
\author{T. Kessler}
\affiliation{Institut f\"ur Physik, Johannes Gutenberg-Universit\"at, Staudinger Weg 7, 55128 Mainz, Germany}
\author{M. Kl\"aui}
\affiliation{Institut f\"ur Physik, Johannes Gutenberg-Universit\"at, Staudinger Weg 7, 55128 Mainz, Germany}
\author{M. Jourdan}
 \email{Jourdan@uni-mainz.de}
\affiliation{Institut f\"ur Physik, Johannes Gutenberg-Universit\"at, Staudinger Weg 7, 55128 Mainz, Germany}

\date{\today}

\begin{abstract}
Many Heusler compounds are predicted to be ferromagnetic half metals in the bulk, which makes them promising compounds for spintronics. However, for devices the transport spin polarization at specific interfaces requires optimization. We show that investigations of the unidirectional magnetoresistance provide an alternative approach to access this quantity. Based on a Wheatstone-bridge design we probed the unidirectional magnetoresistance of Co$_2$MnSi/(Ag, Cu, or Cr)($0.5$~nm)/Pt (or Ta) multilayers and separate the spin-dependent unidirectional spin Hall magnetoresistance from other contributions. We demonstrated that by the insertion of a thin epitaxial Ag layer the spin-dependent contribution is doubled corresponding to a significant increase of the transport spin polarization, which is discussed in the framework of highly spin polarized interface states.   
\end{abstract}

\maketitle

\section{INTRODUCTION}
The unidirectional spin Hall magnetoresistance (USMR) in ferromagnetic metal (FM) / heavy metal (HM) bilayers depends on the direction, i.\,e.\,forward or backward, of the probing current and is switched by reversal of the magnetization direction of the FM layer \cite{Avc15, Ole15}. This is explained by the current direction dependent spin polarization created at the FM/HM-interface by the spin Hall effect \cite{Dya71, Hir99, Kat04,Kim07} (and/or Rashba-Edelstein effect), separating electrons with different spin orientations perpendicular to the current direction. Due to spin-orbit coupling, which is large in HMs such as Platinum, a current in the HM generates a transport spin polarization which scales with the applied current density \cite{Sin15,Sta17}. These spin polarized charge carriers at least partially enter the FM in which, depending on its magnetization direction, they scatter either into the majority or minority electron states. This results in a contribution to the resistivity similar to the current-in-plane (CIP) giant magnetoresistance (GMR) \cite{Bai88, Bin89}. A corresponding theoretical model of the USMR was recently developed by Zhang et al. \cite{Zha16}, within which the electron mobility depends on the spin orientation in the FM. Another contribution to the USMR (and GMR) originates from interface resistance, i.\,e.\,a spin orientation dependent reflection/ transmission coefficient \cite{Avc15, Yas16, Avc18}, which is relevant for GMR devices as well. Additionally, also electron-magnon spin-flip scattering \cite{Dem11} via spin-disorder resistivity of the FM \cite{Kas56,Yas16,Lan16,Li17,Avc18} and pure thermal effects as discussed below can generate an unidirectional magnetoresistance (UMR).

The apparent similarity of the USMR and of the CIP-GMR motivates the use of the rather simple USMR geometry for investigations of the transport spin polarization of specific materials. Here we present the optimization of Heusler-compound/normal metal interfaces as an example of our approach of GMR device optimization.

Potentially halfmetallic, i.\,e.\,100\% spin polarized, Heusler compounds \cite{Gra11} are promising as FM layers in GMR devices. Indeed, after extensive materials optimization processes huge current-perpendicular-plane  GMR values up to 184\% at low temperatures \cite{Sak12} and 63\% at room temperature \cite{Kub18} were obtained using Heusler electrodes. However, no corresponding large current-in-plane GMR was reported up to now. In this framework the magnitude of the USMR can be used to compare the transport spin polarisation of different Heusler interfaces in a simple CIP geometry.

Specifically we investigate the USMR of Co$_2$MnSi(001)/(Ag, Cu, or Cr)($0.5$~nm)/Pt multilayers. For this Heusler compound we previously identified close to 100\% spin polarisation of the free Co$_2$MnSi(001) vacuum surface by UV-photoemission spectroscopy, which we identified to be related to a highly spin polarized surface resonance \cite{Jou14, Bra15}. High energy X-ray photoemission spectroscopy (HAXPES) of different buried Co$_2$MnSi(001)/metal interfaces provided evidence that this surface resonance is preserved in contact with Ag(001) layers, but diminished or suppressed by other metals \cite{Lid18}.

To enable USMR based conclusions concerning transport spin polarization at the interfaces described above, some pitfalls of unidirectional magnetoresistance measurements have to be discussed and investigated: the USMR is a small effect with a relative resistance change of $\simeq 3\cdot10^{-5}$ at current densities of $j\simeq 10^7$~Acm$^{-2}$ only \cite{Avc15,Ole15,Lv18}. As it needs to be extracted from magnetic field dependent hysteresis loops or directional dependent measurements of the magnetoresistance, the influence of other magnetoresistance effects and thermal drifts on the experimental results has to be considered. In general, anisotropic magnetoresistance (AMR) \cite{Cam70} and ordinary spin Hall magnetoresistance (SMR) \cite{Nak13,Hah13,Alt13} effects are much larger than the USMR. The SMR appears also at interfaces between ferromagnetic insulators and HMs and originates from a magnetization dependent backscattering of the spin-polarized electrons. Although it is related to the USMR, it can be separated from it by its different angular dependance \cite{Mia14}.
 
Also thermal effects potentially generate an unidirectional magnetoresistance, as the in-plane current results in a heat gradient perpendicular to the layers. This heat gradient  results in a lateral voltage e.\,g.\,due to the anomalous Nernst effect (ANE) \cite{Kik13, Avc14}, which does not reverse sign when the current direction is changed. Thus it result qualitatively in the same current and field direction dependence as the USMR, making their separation difficult. In general, this and other effects like the longitudinal spin Seebeck effect (LSSE) \cite{Uch14} merge into an unidirectional magnetoresistance (UMR). Based on refs.\,\cite{Avc15,Zha16} we define the UMR as 
\begin{equation}
UMR(j)=\frac{\rho_{xx}(j^{HM})-\rho_{xx}(-j^{HM})}{\frac{1}{2}(\rho_{xx}(j^{HM})+\rho_{xx}(-j^{HM}))}
\end{equation}
with the specific longitudinal resistivity $\rho_{xx}$ of the FM/HM bilayer, which due to the UMR depends on the current density $j^{HM}$ in the heavy metal layer generating the spin polarized current.

\section{EXPERIMENTAL TECHNIQUES}
Our experiments are based on epitaxial thin films of the Heusler compound Co$_2$MnSi, which were deposited by radio frequency magnetron sputtering from a single stoichiometic target. The films were deposited on MgO(100) substrates at room temperature and subsequently annealed in-situ in ultra high vacuum conditions ($\simeq 10^{-9}$~mbar) at a temperature of 550~$^{\rm o}$C. X-ray and in-situ reflection high-energy electron diffraction (RHEED) demonstrated that the Heusler thin films show L2$_{\rm 1}$ ordering in the bulk as well as at the surface \cite{Jou14,Fin15}. Consistent with RHEED patterns in-situ scanning tunneling microscopy of the Heusler thin films shows in general atomically flat terraces \cite{Her09}. To obtain the USMR multilayer structures all other layers were deposited in-situ by rf sputtering as well. Patterning of the samples was performed by standard optical lithography in combination with Ar ion beam etching.

Up to now most published USMR investigations were based on measurements of the second harmonic voltage generated by an ac current through a single conductor stripe \cite{Avc15,Ole15,Avc18}, but also dc measurements were performed on topological insulators \cite{Yas16}.  We chose a Wheatstone bridge \cite{Mei09} like patterning of our samples to enable dc measurements of the UMR in the presence of a large AMR effect of the Heusler compound. A schematic representation of the bridge is shown in Fig.\,1.
\begin{figure}[htb]
\includegraphics[width=1.0\linewidth]{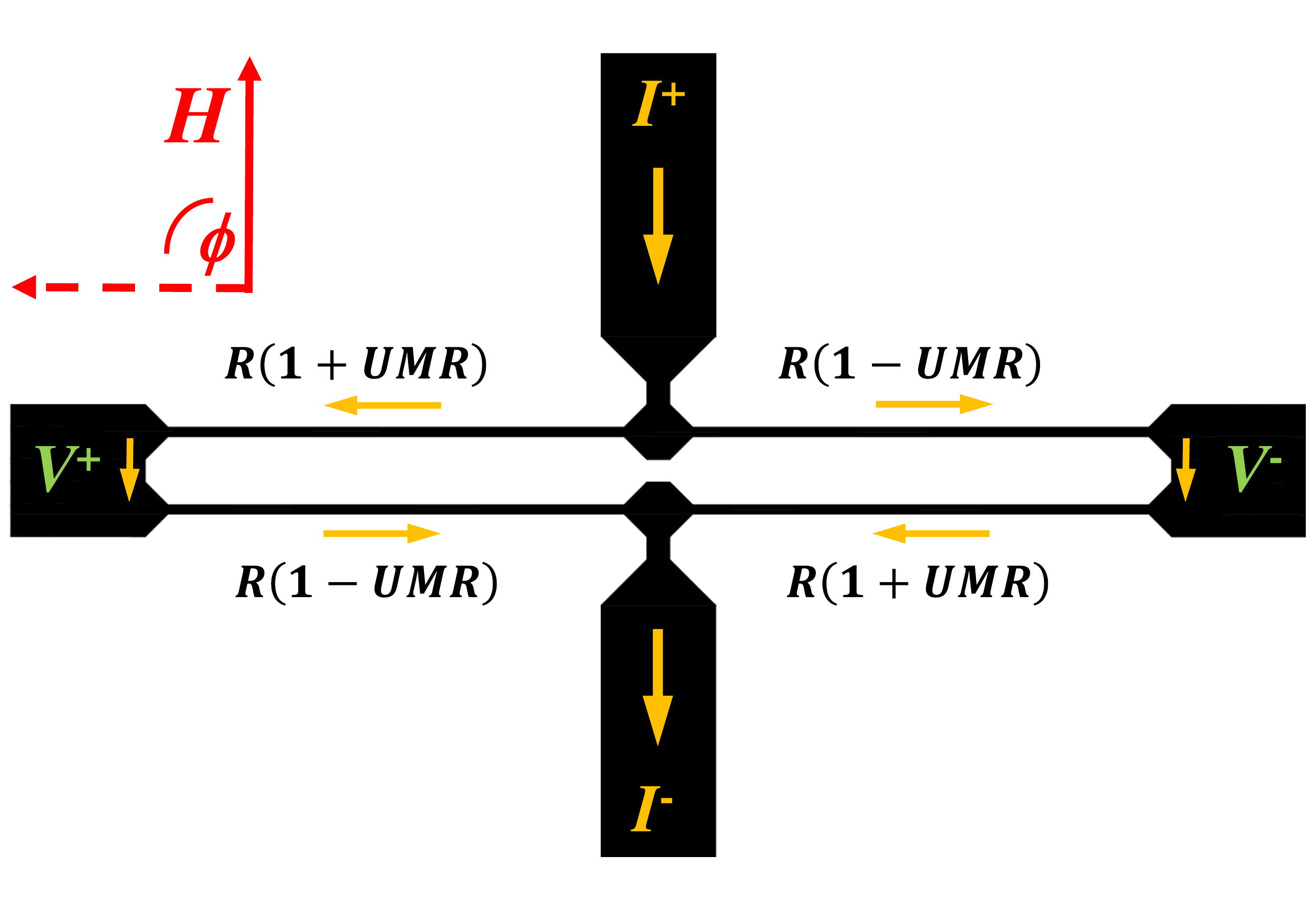}
\caption{Schematic of the Wheatstone bridge sample layout for measurements of the UMR. Each leg of the bridge has a length of $2400$~${\rm \mu}$m and a width of 6~${\rm \mu}$m.}
\end{figure}
This is possible because only the UMR generated dc voltages is directly measured, which originates from tiny relative changes of the large resistance of the legs of the bridge. Additionally, the bridge design compensates the effects of drifts of the background temperature during the measurement. However, a small AMR originating from the voltage contact pads, as well a negligible ordinary Hall contribution due to misaligned magnetic field contribute to the voltage across the bridge.
 
With the design shown in Fig.\,1 the bridge is fully compensated and unidirectional contributions to the resistance of the four legs of the bridge generate a difference of the potentials labeled $V^+$ and $V^-$. The UMR effect detunes pairwise the resistances of the legs in opposite direction as shown in Fig.\,1 resulting in a voltage $V=V^+-V^-$proportional to the detuning $\pm R\cdot UMR$, i.\,e.\,
\begin{equation}
UMR(I)=\frac{V^+-V^-}{I\cdot R}
\end{equation}
with the current $I$ send through the bridge and the resistance $R$ of a single leg of the bridge.

A dc current $I$ was send continuously through the bridge by a Keithley 224 current source, while it was rotated by a stepper motor in an external in-plane field of $\mu_0 H=75$~mT. The voltage $V=V^+-V^-$ was measured using a Keithley 2182 Nanovoltmeter.  

Fig.\,2 shows an example of a measurement of the Wheatstone-bridge voltage $V=V^+-V_-$ normalized to the current density $j=1.67\times10^6$~A/cm$^2$ flowing through the bridge. According to Eq.\,2 this corresponds to a measurement of the absolute resistance change UMR~$\times$~R, which is plotted versus the angle $\phi$ (see inset of Fig.\,1) between the applied in-plane magnetic field $\lvert H \rvert = 75$~mT and the direction of the legs of the bridge. 
 
\begin{figure}[htb]
\includegraphics[width=1.0\linewidth]{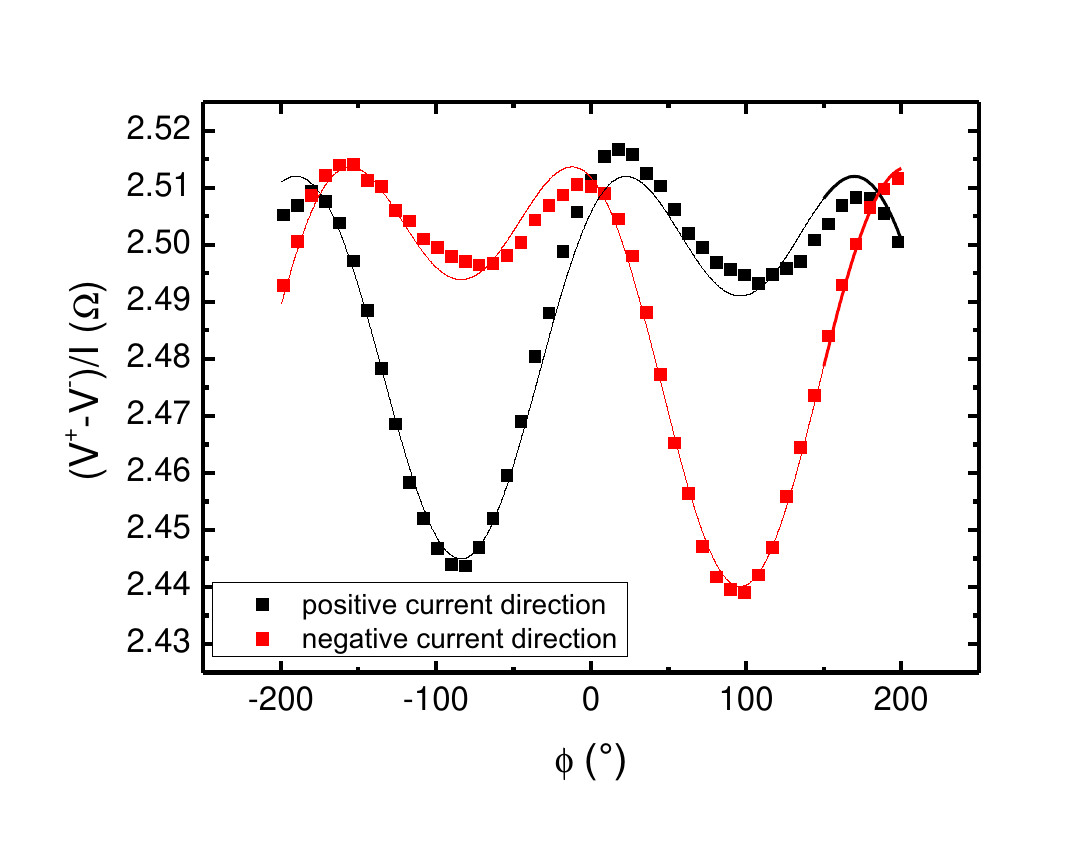}
\caption{Field direction dependence of the voltage across the Wheatstone bridge shown for both current polarities. The fits (red and black lines) of the angular dependance include UMR and residual AMR contributions (see main text).}
\end{figure}

The angular dependance can be fitted by a sum of the antisymmetric contribution of the UMR \cite{Avc15,Avc18} and of residual symmetric contributions presumably due to the AMR \cite{Mcg75} of the contact pads: 
\begin{equation}
\frac{V^+-V^-}{I}=R_0+R_{AMR}\cdot \cos^2\phi\pm UMR \cdot R\cdot\sin\phi
\end{equation}
The sign of the UMR depends on the current direction through the Wheatstone bridge and can be positive or negative, respectively. By taking the voltage difference of measurements with opposite current polarity the residual symmetric contributions cancel each other and only unidirectional resistance contributions remain (see inset of Fig.\,3). The main panel of Fig.\,3 shows the resulting current density dependance of the UMR.
     
\begin{figure}[htb]
\includegraphics[width=1.0\linewidth]{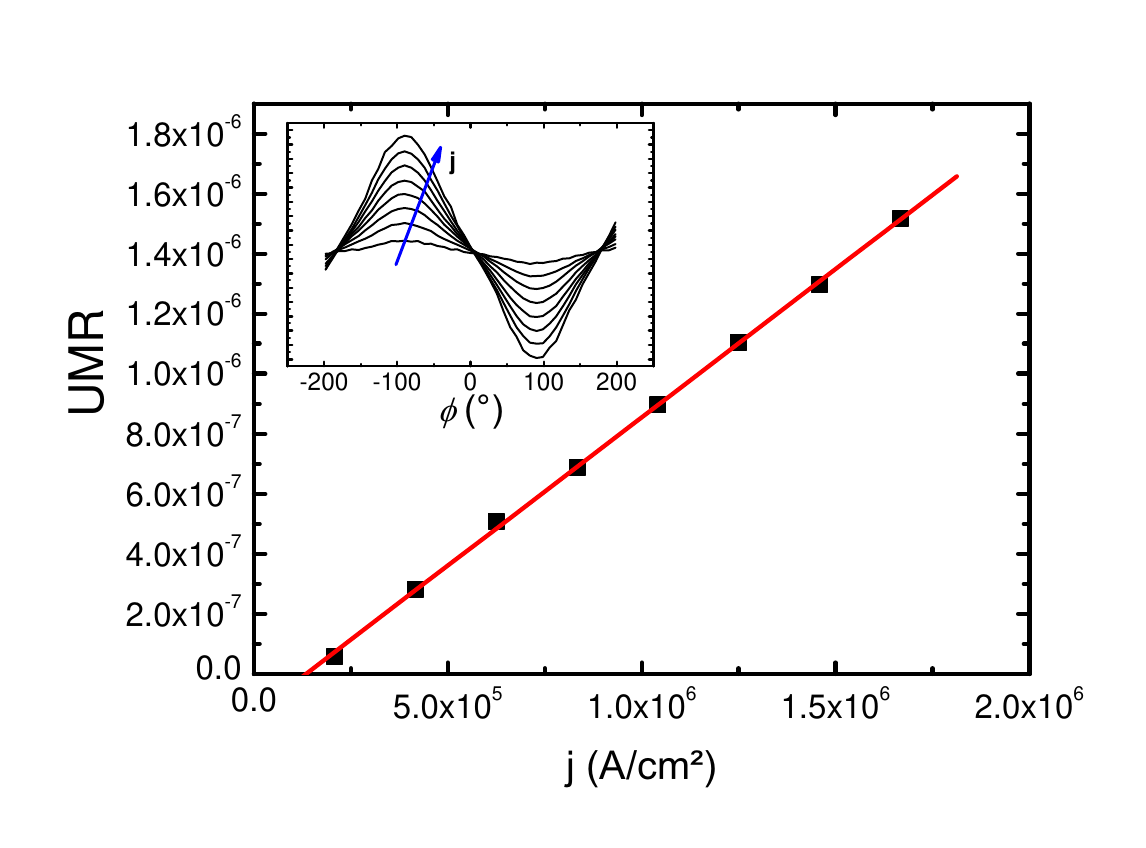}
\caption{Dependance of the UMR on the current density in the legs of the Wheatstone-bridge. The inset shows the angular dependent UMR for the all current densities corresponding to the data points of the main panel.}
\end{figure}

\section{ORIGIN OF THE UMR}
The observed linear dependence of the UMR on the current density is consistent with a spin-dependent USMR as well as with an ANE/PSSE based origin. Within the framework of the former effect the accumulated spin polarization driven by the current in the Pt layer scales with the current density \cite{Zha16}. However, the latter effects result from a thermal gradient perpendicular to the bilayer system, which also should increase with increasing current density due the ohmic heating within the Pt and Co$_2$MnSi layers. We separate these effects by studying the dependence of the UMR on the thickness of the Co$_2$MnSi layers: whereas the thermal contribution is expected to grow with increasing thickness of the ferromagnetic layer \cite{Yin17}, the thickness dependence of the USMR was predicted to show a clear maximum at the spin diffusion length of the Heusler compound \cite{Zha16}.

Experimentally, as shown in Fig.\,4, in the case of the Co$_2$MnSi/Pt samples we observed a shallow maximum of the UMR for a Co$_2$MnSi layer thickness of $\simeq 7$~nm, which is similar to the optimum thickness of the ferromagnetic layer ($\sim 5$~nm) of Co/Pt bilayers reported by \cite{Yin17}. For Heusler layer thicknesses larger 10~nm the UMR increases qualitatively in the same way as calculated for a thermally generated ANE signal \cite{Yin17}. In our case the thermal contribution is relatively large, which might be related to the recently reported exceptionally large ANE of Co$_2$MnAl$_{1-x}$Si$_x$ \cite{Sak18}. Nevertheless, the USMR, in which we are primarily interested, is dominating the UMR for thin layers.      
   
\begin{figure}[htb]
\includegraphics[width=1.0\linewidth]{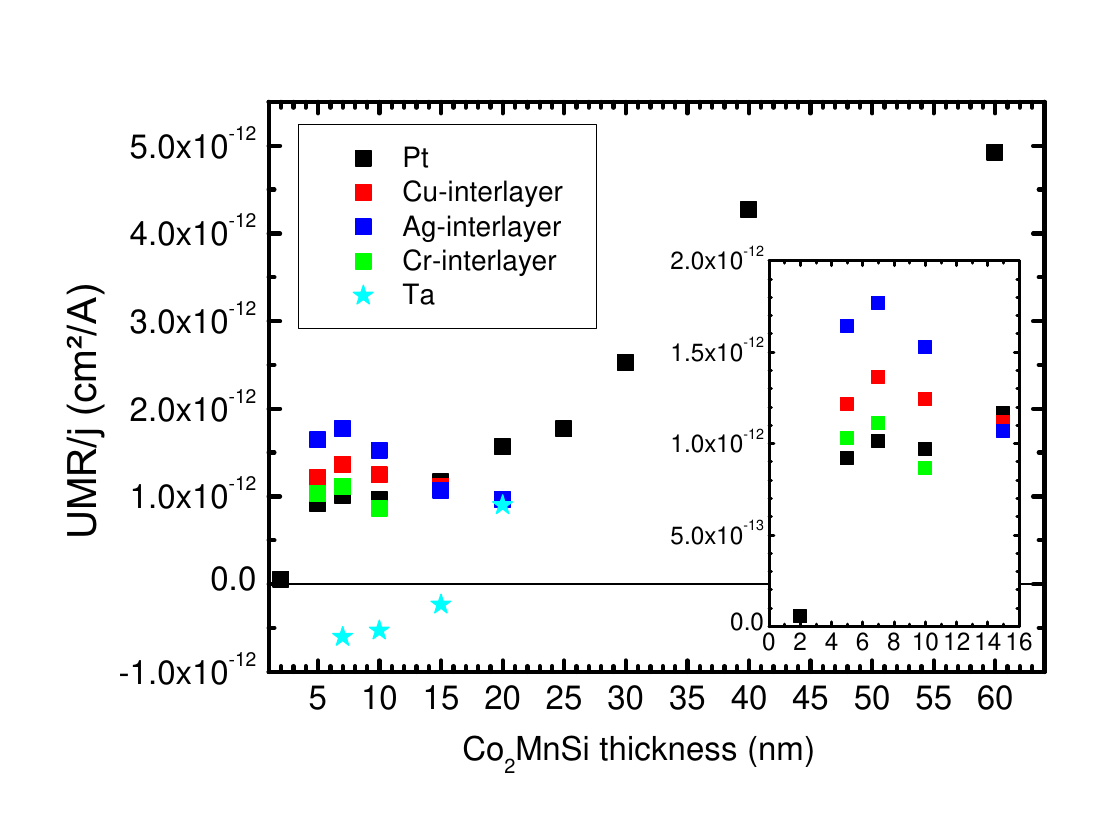}
\caption{Normalized UMR of Co$_2$MnSi/interlayer(0.5nm)/Pt(5nm) multilayers versus thickness of the Co$_2$MnSi layer. Black squares: Co$_2$MnSi/Pt(5nm) without interlayer; red/ blue/ green squares: Co$_2$MnSi/Cu, or Ag, or Cr (0.5nm)/Pt(5nm); cyan stars: Co$_2$MnSi/Ta(5nm) without interlayer. The inset enlarges the Co$_2$MnSi thickness range dominated by the spin-dependent USMR.}
\end{figure}

Further evidence for this conclusion is obtained by replacing the HM Pt by Ta with the opposite sign of it's spin Hall angle ($\Theta_{SH}^{Pt}=0.1, \Theta_{SH}^{Ta}=-0.07 $ \cite{Wan14}). As the sign of the thermal gradient discussed above is not changed by this, the USMR contribution can now be separated from the thermal contributions to the UMR. Additionally, also the spin-disorder contribution to the UMR discussed in the introduction is always positive \cite{Avc18}. In contrast to the spin-dependent USMR contributions relevant for the GMR effect it does not change sign with the spin Hall angle of the HM.

Indeed, as shown in Fig.\,4, we obtained a positive UMR for thin Co$_2$MnSi($\simeq 10$nm)/Pt bilayers and a negative UMR for Co$_2$MnSi($\simeq 10$nm)/Ta bilayers. This provides strong evidence for the spin-dependent USMR dominating the total UMR in  thickness range of the Heusler thin films. For a larger thicknesses of the Ta layer (20 ~nm) again a positive UMR was measured, which is consistent with the assumption of increasing positive thermal contributions for thicker Heusler layers.  

\section{TRANSPORT SPIN POLARIZATION}
Having established the thickness range in which the UMR of our samples is dominated by the spin-dependent USMR effect we can now investigate the influence of thin metallic interlayers between the FM and the HM on the interface spin polarization. Based on our previous investigations of the surface/interface states of Co$_2$MnSi by photoemission spectroscopy \cite{Lid18} we expect that the specific choice of the metallic partner material at the interface has a strong influence on the interface states of the Heusler compound, which should influence the magnitude of the spin-dependent USMR. Thus we investigated the UMR of samples with additional layers of polycrystalline Cu, epitaxial Cr(001), and epitaxial Ag(001), each with a thickness of $0.5$~nm at the interface between Co$_2$MnSi(001) and Pt (polycrystalline). As shown in Fig.\,4 these interface layers modify the UMR only in the thickness range of Co$_2$MnSi, which we associated with the spin-dependent USMR providing further strong evidence for the validity of this assumption. Thus we are now able to discuss the effects of the interlayer insertion in the framework of transport spin polarization as reflected by the spin-dependent USMR.

The USMR obtained with the inserted epitaxial Ag(001) layer is about twice as large as the value obtained without any inserted layer. On the other hand, insertion of an epitaxial Cr(001) layer results in no modification of the spin-dependent USMR compared to directly interfacing with polycrystalline Pt. Using polycrystalline Cu as an interlayer raises the USMR, but much less than the Ag(001) layer.

This is fully consistent with our previous spin-integrated high energy x-ray photoemission spectroscopy (HAXPES) experiments on Co$_2$MnSi(001)/metal(2nm)/AlO$_x$(2nm) thin films samples, which allowed us to associate a shoulder feature in the photoemission intensity with a highly spin polarized interface state based on band structure calculations \cite{Lid18}. The distinctiveness of this HAXPES feature near the Fermi energy is correlating perfectly with the relative magnitude of the spin-dependent USMR of the samples with different metallic interlayers discussed here. 

We conclude that measurements of the USMR are a GMR device independent  method for the comparison of transport spin polarizations of samples with modified interfaces. Although the Heusler compound Co$_2$MnSi is a bulk half metal with a highly spin polarized resonance at the Co$_2$MnSi(001) surface \cite{Jou14, Bra15}, the spin polarization can be strongly diminished at interfaces with other metals. Our UMR investigations in the regime dominated by the spin-dependent USMR show, that the transport spin polarization relevant for e.\,g.\,GMR devices is strongly reduced at interfaces with polycrystalline Pt and epitaxial Cr, whereas it is relatively large at interfaces with epitaxial Ag. Although we discuss transport experiments here, this might be explained in analogy to our previous photoemission spectroscopy experiments by the destruction or shifting of highly spin polarized interface states \cite{Lid18}. The observed spin-dependent USMR is consistent with reports of large current perpendicular plane (CPP)-GMR values obtained with Heusler electrodes in combination with Ag spacer layers, but not with Cr layers. Our results indicate, that the combination of Co$_2$MnSi and Ag layers should also enable large current in plane (CIP)-GMR values.

\section {SUMMARY}
In summary, using a Wheatstone-bridge geometry the unidirectional magnetoresistance (UMR) of halfmetallic Heusler compound/Pt bilayers can be obtained by dc measurements. Although the anomalous Nernst effect of Heusler compounds is relatively large, this thermal contribution to the unidirectional magnetoresistance of Co$_2$MnSi/Pt bilayers becomes relevant for thick layers of the Heusler compound only. We were able to define a thickness regime of the Heusler thin films in which the spin-dependent unidirectional spin Hall magnetoresistance (USMR) dominates the total UMR. Within this regime, i.\,e.\,focusing on thin Heusler layers ($\simeq 7$~nm) the USMR was used to probe the influence of $0.5$~nm Ag, Cu and Cr layers at the Co$_2$MnSi/Pt interface on the transport spin polarization. We demonstrated that the insertion of a thin epitaxial Ag(001) layer clearly increases this quantity compared the direct interface with polycrystalline Pt as well as with the other interlayer materials. This result is consistent the supression or shifting of spin polarised interface states as discussed previously \cite{Lid18}. Thus interface tuning promises an improved performance of Heusler based spintronics specifically including CIP-GMR devices. Investigations of the spin-dependent USMR provide an alternative approach for probing such effects without the technically challenging actual preparation of GMR devices.

{\bf ACKNOWKEDGEMENTS}

Financial support by the the German Research Foundation (DFG) via project Jo404/9-1 is acknowledged.


\end{document}